\documentclass[sigconf]{acmart}

\usepackage[table]{xcolor}
\definecolor{rowgray}{gray}{0.93}

\AtBeginDocument{%
  }

\copyrightyear{2026}
\acmYear{2026}
\setcopyright{cc}
\setcctype{by}
\acmConference[ICSE-FoSE]{2026 IEEE/ACM 48th International Conference on Software Engineering: Future of Software Engineering }{April 12--18, 2026}{Rio de Janeiro, Brazil}
\acmBooktitle{2026 IEEE/ACM 48th International Conference on Software Engineering: Future of Software Engineering (ICSE-FoSE), April 12--18, 2026, Rio de Janeiro, Brazil}
\acmPrice{}
\acmDOI{10.1145/3793657.3793884}
\acmISBN{979-8-4007-2479-4/2026/04}

\begin{document}

\title{When Code Becomes Abundant: Redefining Software Engineering Around Orchestration and Verification}


\author{Karina Kohl}
\affiliation{%
  \institution{Institute of Informatics, UFRGS}
  \city{Porto Alegre}
  \country{Brazil}}
\email{karina.kohl@inf.ufrgs.br}
\orcid{0000-0002-2964-4681}

\author{Luigi Carro}
\affiliation{%
  \institution{Institute of Informatics, UFRGS}
  \city{Porto Alegre}
  \country{Brazil}
}\email{carro@inf.ufrgs.br}
\orcid{0000-0002-7402-4780}

\renewcommand{\shortauthors}{Kohl and Carro}

\begin{abstract}

Software Engineering (SE) faces simultaneous pressure from AI automation (reducing code production costs) and hardware/energy constraints (amplifying failure costs). We position that SE must redefine itself around human discernment—intent articulation, architectural control, and verification—rather than code construction. This shift introduces 'accountability collapse' as a central risk and requires fundamental changes to research priorities, educational curricula, and industrial practices. We argue that Software Engineering, as traditionally defined around code construction and process management, is no longer sufficient. Instead, the discipline must be redefined around intent articulation, architectural control, and systematic verification. This redefinition shifts Software Engineering from a production-oriented field to one centered on human judgment under automation, with profound implications for research, practice, and education.
\end{abstract}

\ccsdesc[500]{Software and its engineering~Software design engineering}

\keywords{Software Engineering, Orchestration, Verification, AI-Generated Code, Architectural Governance, Accountability}


\maketitle

\section{Introduction: The Compression of Software Engineering}

When machines write code faster than humans can read it, Software Engineering (SE) can no longer be primarily about programming. It must become the discipline of deciding what should exist and proving it behaves accordingly.

Software Engineering faces pressure from two directions at once. From above, large language models enable automated code generation, testing, documentation, and deployment artifacts. GitHub Copilot users report 55\% faster task completion rates~\cite{copilot_productivity}, while the marginal cost of code generation approaches levels previously unimaginable~\cite{zero_marginal_cost}. From below, the end of Dennard scaling in the mid-2000s and the slowdown of Moore’s Law—with transistor shrinking now requiring innovations beyond traditional scaling~\cite{moores_law_2017,dennard_scaling_end,end_of_scaling}—combine with rising energy costs and renewed non-functional demands: efficiency and low energy footprint are again valuable properties. Together, these opposing forces erode long-standing assumptions about how software is produced, evolved, and validated.

\begin{figure*}[t]
    \centering
    \includegraphics[width=0.5\linewidth]{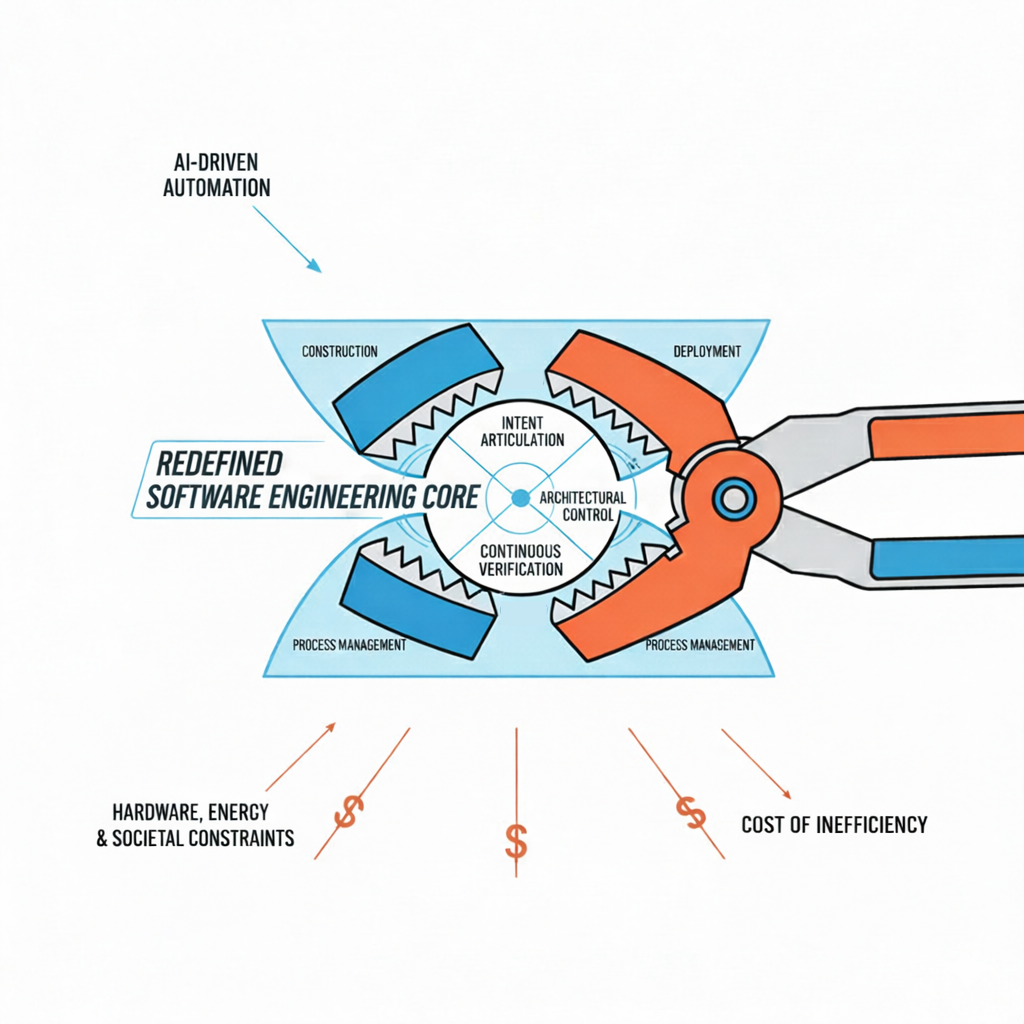}
    \caption{Conceptual illustration of SDLC compression under automation and physical constraints.}
    \label{fig:compression}
\end{figure*}

Figure~\ref{fig:compression} illustrates this \emph{compression} by depicting the traditional Software Development Life Cycle squeezed between these forces. Pressure from above collapses construction, deployment, and routine maintenance by making code generation cheap, fast, and continuous. Pressure from below—imposed by physical limits, rising energy costs, and regulatory constraints—increases the stakes of errors, inefficiencies, and unintended behaviors. Here, \emph{regulatory constraints} are the external rules and accountability mechanisms that make shipping wrong software more costly—legally, financially, and operationally—especially when AI accelerates production, and they increasingly require privacy- and security-by-design, traceability, and supply-chain provenance, and auditable compliance in safety- and mission-critical settings. Under this compression, lifecycle phases are flattened, displaced, or absorbed into automated pipelines. What remains is a redefined core concentrated around three activities: intent articulation, architectural control, and continuous verification.


Consider a financial firm's AI regenerates risk modules weekly, leading to a \$50 million loss. Because the system is reproducible (regenerable from specs) but not explainable, the incident version of the code is lost and causal chains are obscured. This illustrates accountability collapse: the erosion of links between human decisions and system behavior when automated synthesis—rather than manual design—determines software structure.

In this position paper, we argue that Software Engineering must be \emph{redefined rather than merely extended}. When code is abundant, automatically generated, and disposable, a discipline organized around construction, prescriptive methodologies, and manual maintenance becomes insufficient. What remains scarce is not implementation capacity, but human discernment: the ability to decide what should be built, to govern how systems are generated, and to continuously verify—against evolving intent, physical and logical constraints, and societal values—that systems behave as intended. Preventing accountability collapse, we contend, is the defining challenge of Software Engineering under compression.

\section{The Reconfiguration of the Traditional SDLC}
The canonical Software Development Life Cycle distributes human effort across requirements, design, implementation, testing, deployment, and maintenance. Under pervasive AI-driven automation, this decomposition no longer reflects where human judgment is most critical. As automated systems absorb construction and operational tasks, traditional phase boundaries lose explanatory power and practical relevance. This reconfiguration is not merely organizational—it directly enables accountability collapse by obscuring where intent is defined, where decisions are made, and where responsibility for system behavior can be meaningfully assigned.

\subsection{Reorganizing Responsibilities under Compression}
AI does not eliminate work; it relocates the primary bottleneck from production to human discernment. Under compression, the SDLC reorganizes around two dominant loci of human responsibility:

\begin{itemize}
\item \textbf{Orchestration:} expressing goals, constraints, values, and non-functional requirements in forms that meaningfully guide automated synthesis. This includes requirements elicitation, architectural decisions, policy enforcement for autonomous systems, and explicit specification of trade-offs among competing objectives.

\item \textbf{Verification:} continuously evaluating whether generated systems—often opaque, evolving, and partially autonomous, faithfully realize intent without unacceptable side effects. This includes validation of generated artifacts, testing as executable specification, runtime monitoring, and review-time human oversight.
\end{itemize}

Orchestration and verification are not new lifecycle stages replacing existing activities. Rather, they represent conceptual poles around which traditional Software Engineering responsibilities reorganize. Established concerns—requirements, architecture, testing, security, and maintenance—are not eliminated. but are reinterpreted according to whether they primarily constrain automated generation (orchestration) or evaluate its outcomes (verification).

Table~\ref{tab:sdlc_reconfiguration} illustrates this reorganization. Requirements shift from upfront specification documents to continuous intent modeling because AI systems require ongoing constraint specification rather than static artifacts~\cite{intent_specifications}. Architecture transitions from design guidance to a control surface because it defines the boundaries within which automated synthesis operates. Testing becomes verification rather than downstream quality assurance because generated code cannot be trusted through inspection alone—executable tests serve as the primary mechanism for establishing trust in opaque artifacts. Maintenance transforms from incremental bug fixing to continuous verification across regenerations because systems may be restructured entirely rather than patched. Security becomes both a generation constraint (what must not be synthesized) and adversarial validation (what vulnerabilities emerged despite constraints).

In the following sections, we examine two reinterpretations in detail: architecture as the primary control for orchestration (Section \ref{arch}), and maintenance as a longitudinal verification challenge under repeated regeneration (Section \ref{maintanance}). These examples demonstrate how orchestration and verification materialize in practice.

\begin{table*}[t]
\centering
\caption{Shift in the locus of human discernment across Software Engineering responsibilities under SDLC compression}
\label{tab:sdlc_reconfiguration}
\rowcolors{2}{white}{rowgray}
\begin{tabular}{p{2.5cm} p{5cm} p{9cm}}
\toprule
\textbf{SE Responsibility} &
\textbf{Traditional SDLC Role} &
\textbf{Role under Compression} \\
\midrule
Requirements &
Upfront specification phase &
Continuous intent modeling and constraint articulation (Orchestration) \\

Architecture &
Design artifact guiding implementation &
Control surface constraining automated generation and enabling traceability (Orchestration) \\

Testing &
Downstream quality assurance &
Executable specification and primary mechanism of trust (Verification) \\

Maintenance &
Bug fixing and incremental evolution &
Continuous verification across regenerations and evolving intent (Verification) \\

Security &
Specialized activity or review stage &
Constraint on generation and adversarial validation throughout the lifecycle (Orchestration \& Verification) \\

Human Oversight &
Managerial or supervisory role &
Active governance, rejection, and intervention in autonomous and semi-autonomous systems (Orchestration \& Verification) \\
\bottomrule
\end{tabular}
\end{table*}

\subsection{Architecture as a Control Surface, Not a Design Artifact} \label{arch}

In the compressed lifecycle, architecture is the primary mechanism through which orchestration is exercised and rendered verifiable. Architecture functions not merely as a technical abstraction, but as a form of \emph{governance}. By constraining how systems are decomposed, generated, and composed, architectural decisions establish boundaries within which automated synthesis operates. These boundaries are essential for preventing accountability collapse: they preserve traceability between intent, generation, and observed behavior, and define where responsibility can be meaningfully located.

Under these conditions, poor architecture is no longer merely a source of technical debt—it constitutes a loss of human agency. Systems with inadequate architectural constraints become increasingly difficult to reason about, constrain, or audit as generation and regeneration cycles accumulate. Conversely, well-chosen architectural decompositions act as scaffolding for both automated generation and verification. They restrict the space of possible behaviors, enable targeted validation, and preserve meaningful intervention points despite overwhelming automation.

As automation absorbs construction tasks, human discernment emerges as the new scarce resource. Software Engineering work is defined less by production capacity and more by the ability to manage cognitive load, calibrate trust in AI agents, detect silent failures and hallucinations, identify misaligned intent, and exercise ethical responsibility over system behavior. In this sense, Software Engineering shifts from a discipline of construction to one of \emph{sensemaking under automation}, where humans are responsible not for writing code, but for understanding, constraining, and legitimizing what machines produce.

This shift is not speculative. Empirical signals from the community reinforce it: surveys of senior researchers increasingly emphasize greater—not lesser—responsibility for humans in AI-assisted development workflows~\cite{fose_2026}. As a result, Software Engineering's core contribution is no longer efficiency in building software, but sustained human oversight of systems that would otherwise exceed our capacity to comprehend and control.

\subsection{Maintenance Is Not Simpler—It Is More Dangerous}\label{maintanance}
A prevailing assumption in AI-driven software development is that maintenance becomes trivial once systems can be regenerated on demand. This assumption is deeply misleading. Under SDLC compression, maintenance is no longer a downstream phase focused on code repair, but rather the longitudinal extension of verification across successive regenerations.

Regeneration presupposes stable intent, complete specifications, and reproducible behavior—conditions that rarely hold in complex socio-technical systems where requirements evolve, contexts shift, and values are negotiated rather than fixed. In practice, regeneration amplifies the risk of accountability collapse. When systems are repeatedly regenerated rather than incrementally evolved, historical responsibility becomes blurred, causal explanations for failures are obscured, and the connection between design decisions and observed behavior weakens. Maintainable systems are replaced by artifacts that are \emph{reproducible} yet poorly \emph{understood}, making it increasingly difficult to answer fundamental questions: Why does this system behave as it does? Who is responsible when it fails?

Consider the financial risk assessment scenario from Section~1. After the \$50 million loss, the development team can regenerate the exact system configuration from the day of the incident—but the architectural constraints, generation prompts, and human override decisions that shaped that configuration have been overwritten across dozens of subsequent iterations. The system is auditable in a snapshot sense, but its \emph{evolution} is opaque. This is maintenance under compression: not a problem of fixing bugs, but of maintaining accountability across a succession of generated artifacts.

Under this redefinition, maintenance does not disappear; it is fundamentally reframed as a continuous verification process. The central challenge becomes ensuring that successive regenerations remain aligned with evolving intent, constraints, and societal expectations. Success requires new practices: maintaining intent logs alongside code repositories, architectural review gates before and after automated synthesis, and dedicated roles for "regeneration auditors" who trace decisions across system evolution. Without such mechanisms, the convenience of regeneration risks becoming a liability—systems that can be rebuilt at will, but whose behavior cannot be explained or justified.

\section{Implications: A Redefined Discipline}
\subsection{Research}

Software Engineering research must move beyond accelerating production and instead focus on sustaining human discernment in the presence of large-scale automation. Critical research challenges include:

\begin{itemize}
\item How can intent be formally specified to constrain AI synthesis while adapting to evolving requirements?
\item What architectural patterns enable controllability and traceability in systems where components are generated rather than manually implemented?
\item How can we detect misalignment between specified intent and emergent behavior in partially autonomous systems?
\item What verification techniques scale to continuously regenerated systems whose structure changes across iterations?
\item How should responsibility be allocated in hybrid human-AI development workflows to preserve accountability?
\item What metrics capture "maintainability" when maintenance means verifying regeneration rather than patching code?
\end{itemize}

Collectively, these directions address preventing accountability collapse in systems whose behavior is increasingly shaped by automated synthesis.

\subsection{Education}
Educational programs must be reoriented accordingly. Proficiency in languages, frameworks, and tooling—while still useful—can no longer define core competence. Instead, curricula should emphasize: 
\begin{itemize} 
\item \textbf{Architectural reasoning as governance}: teaching students to design systems that constrain and audit automated generation, not merely organize human-written code \item \textbf{Testing as executable specification}: positioning tests as the primary trust mechanism for opaque artifacts rather than post-hoc quality assurance 
\item \textbf{Formal and semi-formal verification}: equipping engineers to establish behavioral guarantees for systems they did not directly construct 
\item \textbf{Ethical responsibility in oversight}: judgment about when to reject, override, or intervene in automated synthesis 
\end{itemize} 
The goal is not to train faster programmers, but to develop engineers capable of reasoning about, constraining, and legitimizing machine-generated software.

\subsection{Practice}

In practice, organizational success will be measured not only by development speed and output volume, but increasingly by the ability to \emph{reject} incorrect, unsafe, or misaligned automatically generated solutions. This requires new practices: intent documentation as a first-class artifact, architectural review integrated into generation pipelines, continuous verification workflows that span regenerations, and explicit accountability mechanisms that trace decisions across system evolution. While these practices impose upfront costs, they prevent the far greater expense of deploying systems whose failures cannot be explained or attributed—a cost amplified by the physical, energy, and societal constraints described in Section~1.

\section{Conclusion: Software Engineering as a Discernment Discipline}

Software Engineering is not disappearing—it is becoming sharper and more demanding. As automation absorbs construction, the discipline's legitimacy rests on its ability to articulate intent, enforce constraints, and establish trust in systems humans no longer directly build. If software can be written by machines, then Software Engineering becomes the discipline of deciding what should exist, what must not, and how we verify the difference.

The future of SE is not faster coding, but rather a detailed,  more
deliberate judgment while undergoing the unprecedented pressure from
automation. Under compression, Software Engineering’s central
contribution is no longer only efficient production, but also the
prevention of accountability collapse in automated systems, whose
scale and complexity would otherwise exceed human comprehension.


\begin{acks}
 Study was financed in part by the Coordenação de Aperfeiçoamento de Pessoal de Nível Superior - Brasil (CAPES) - Finance Code 001.
\end{acks}

\appendix

\section{Appendix}
\subsection{Prompts}

Figure~\ref{fig:compression} was generated using Google Gemini 3 with the following prompt:\textit{"can you create an image for this? The traditional Software Development Life Cycle is compressed by opposing forces. Pressure from above (AI-driven automation) collapses construction, deployment, and process management. Pressure from below (hardware, energy, and societal constraints) amplifies the cost of inefficiency and failure. What remains is a redefined core of Software Engineering centered on intent articulation, architectural control, and continuous verification."}

After generation, some labels in the image contained spelling errors. We therefore used ChatGPT 5.2 to correct the wording while preserving the original visual structure, using the following prompt:\textit{"can you fix this image? the words are spelled wrong"}

\bibliographystyle{ACM-Reference-Format}
\bibliography{sample-base}

\end{document}